\documentclass[preprintnumbers,twocolumn]{revtex4}

\usepackage{graphicx}
\usepackage{epsf}
\usepackage{amsmath,amssymb}
\usepackage{color}

\begin{document}
\preprint{KUNS-2768, NITEP 25}
\title{Microscopic calculation of  inelastic proton scattering off
$^{18}$O, $^{10}$Be, $^{12}$Be,  and $^{16}$C for study of
neutron excitation in neutron-rich nuclei}

\author{Yoshiko Kanada-En'yo}
\affiliation{Department of Physics, Kyoto University, Kyoto 606-8502, Japan}
\author{Kazuyuki Ogata} 
\affiliation{Research Center for Nuclear Physics (RCNP), Osaka University,
  Ibaraki 567-0047, Japan}
\affiliation{Department of Physics, Osaka City University, Osaka 558-8585,
  Japan}
\affiliation{
Nambu Yoichiro Institute of Theoretical and Experimental Physics (NITEP),
   Osaka City University, Osaka 558-8585, Japan}

\begin{abstract}
The microscopic coupled-channel calculation of  inelastic  proton scattering is performed
for the study of neutron excitations in $2^+_1$ states of
$^{18}$O, $^{10}$Be, $^{12}$Be,  and $^{16}$C.
Proton-nucleus potentials in the coupled-channel calculation
are microscopically
derived by folding the Melbourne $g$-matrix $NN$ interaction with
matter and transition densities of target nuclei obtained by the structure model calculation
of antisymmetrized molecular dynamics. The calculated result reasonably reproduces the
elastic and inelastic proton scattering cross sections, and supports the dominant contribution of neutron
in the $2^+_1$ excitation of $^{12}$Be and $^{16}$C as well as $^{18}$O.
Sensitivity of the inelastic scattering cross sections to the neutron transition density is discussed.
The exotic feature of the neutron transition density with the amplitude in the outer region
in $^{12}$Be and $^{16}$C is focused.
\end{abstract}

\maketitle

\section{Introduction}
Shape difference in proton and neutron matter distributions in nuclei
is one of the interesting  phenomena  in unstable nuclei.
To discuss the difference between the neutron and proton deformation
(or collectivity), the neutron and proton transition matrix elements in the ground-band
$2^+_1\to 0^+_1$ transition have been investigated for a long time.
In a naive expectation for ordinary nuclei with the same proton and neutron deformation,
the ratio of the neutron transition matrix element ($M_n$) to the proton one ($M_p$) should be $N/Z$.
However, the relation $M_n/M_p \approx Z/N$ is not satisfied even in stable nuclei
with the proton or neutron shell closure as reported in Ref.~\cite{Bernstein:1981fp}.
For instance, in $^{18}$O and $^{48}$Ca,
the ratio becomes significantly larger than $N/Z$, which
indicates the
neutron dominance in the $2^+_1$ excitation because of the proton shell closure.
The phenomena of the shape difference and/or the neutron dominance have been
suggested also in unstable nuclei such as $^{10}$Be, $^{12}$Be, and $^{16}$C
\cite{Kanada-Enyo:1996zsp,Iwasaki:2000gh,Kanada-Enyo:2004tao,Kanada-Enyo:2004ere,Sagawa:2004ut,Jouanne:2005pb,Takashina:2005bs,Ong:2006rm,Burvenich:2008zz,Takashina:2008zza,Elekes:2008zz,Wiedeking:2008zzb, Yao:2011zza, Forssen:2011dr}.

Experimental information of the proton part $M_p$ can be
directly obtained from the $E2$ strength. By contrast,
determination of the neutron part ($M_n$) is not easy even for stable nuclei.
Instead of direct measurements,
experiments of inelastic hadron scattering have been performed
using such probes as $\alpha$, proton, neutron, and $\pi^-/\pi^+$.
By combining
the hadron scattering data with the electromagnetic data,
$M_n$ and $M_p$ have been discussed based on reaction analysis
(see Refs.~\cite{Bernstein:1981fp,Bernstein:1977wtr} and references therein).
For $^{18}$O, the neutron matrix element of the
$2^+_1\to 0^+_1$ transition has been intensively investigated, and
the anomalously large value of $M_n/M_p\approx 2$ has been reduced from the
inelastic scattering data \cite{Bernstein:1977wtr,Iversen:1978sc,Grabmayr:1980qze,Kelly:1986ysn}
consistently with $B(E2)$ of mirror transitions of $^{18}$Ne and $^{18}$O
\cite{Bernstein:1979zza}.

In study of unstable nuclei,
the neutron collectivity, i.e., the $M_n/M_p$ ratio has been investigated extensively
with the inelastic proton scattering experiments in
inverse kinematics using radioactive ion beam
\cite{Iwasaki:2000gh,Jouanne:2005pb,Ong:2006rm,Jewell:1999jme,Khan:2000rac,Khan:2001gbs,Scheit:2001ys,Becheva:2006zz,Elekes:2008zz,Campbell:2007zz,Elekes:2009zz,Aoi:2010ah,Michimasa:2014qca,Riley:2014vnp,Corsi:2015afa,Cortes:2018izn}.
However, the reaction analysis still contains model ambiguities, for instance, in the 
proton-nucleus optical potentials,
which are phenomenologically adjusted usually
to elastic scattering cross sections but the applicability has not been well tested for inelastic scattering off exotic nuclei.

Recently, triggered by the complete microscopic folding model calculation by the Melbourne group~\cite{Amos:2000,Karataglidis:2007yj}, the microscopic description of proton-nucleus \cite{Min10,Toy13,Toyokawa:2015zxa,Minomo:2017hjl} and $\alpha$-nucleus \cite{Egashira:2014zda,Toyokawa:2015zxa} elastic scattering, without any free adjustable parameter and phenomenological parametrization, has been developed. Very recently, the framework was successfully applied to $\alpha$-nucleus inelastic processes~\cite{Minomo:2016hgc,Minomo:2017hjl,Kanada-Enyo:2019prr,Kanada-Enyo:2019qbp}.
One of the advantages of this approach is that, once reliable densities of target nuclei are given,
there is no adjustable parameter in the reaction part.
As for the structure part,
proton and neutron matter and transition densities are obtained by microscopic structure model calculations,
which describe characteristics of nuclear properties such as the cluster, deformation, and neutron skin structures in target nuclei.

In this paper, we investigate the inelastic proton scattering to the $2^+_1$ states of $^{18}$O, $^{10}$Be, $^{12}$Be, and
$^{16}$C with the coupled-channel (CC) calculations of the microscopic single-folding model using the
Melbourne $g$-matrix effective $NN$ interaction \cite{Amos:2000}. The proton and neutron matter and transition densities
of the target nuclei are calculated with antisymmetrized molecular dynamics (AMD) \cite{KanadaEnyo:1995tb,KanadaEnyo:1995ir,KanadaEn'yo:2012bj}.
As test cases, we first show application to the proton scattering off $Z=N$ nuclei, $^{12}$C and $^{16}$O.
Then, we apply the same method to the proton scattering off $^{18}$O,  $^{10}$Be, $^{12}$Be, and
$^{16}$C. The sensitivity of the $2^+_1$ cross sections to $M_n$ and $M_p$
is analyzed while focusing on the neutron-proton difference in the transition densities in  $^{12}$Be and
$^{16}$C.

The paper is organized as follows. The next section describes the present framework of the microscopic coupled-channel (MCC) calculation
and that of the structure calculations for target nuclei. Results of $^{12}$C and $^{16}$O are shown in Sec.~\ref{sec:results1},
and results and discussions for the $N\ne Z$ case of $^{18}$O,  $^{10}$Be, $^{12}$Be, and $^{16}$C are given in Sec.~\ref{sec:results2}. Finally, a summary is given in Sec.~\ref{sec:summary}.

\section{Method} \label{sec:method}
The present reaction calculation for the
proton scattering is the MCC calculation
of the single-folding model.
As inputs from the structure calculations, the
target densities are calculated with AMD combined with and without the
cluster model of the generator coordinate method (GCM).
The AMD  and AMD+GCM calculations of target nuclei are the same as those of
Refs.~\cite{Kanada-Enyo:1999bsw,Kanada-Enyo:2003fhn,Kanada-Enyo:2004tao,Kanada-Enyo:2019prr,Kanada-Enyo:2019hrm,Kanada-Enyo:2019qbp}.
The definitions of densities and form factors in the structure calculation
are explained in Ref.~\cite{Kanada-Enyo:2019prr}.
For details, the reader is referred to those references.

\subsection{Microscopic coupled-channel calculation}

The diagonal and coupling potentials for the nucleon-nucleus system are microscopically calculated
by folding the Melbourne $g$-matrix $NN$ interaction \cite{Amos:2000} with the target densities described in Sec.~\ref{sec3c}.
The Melbourne $g$ matrix is obtained by solving a Bethe-Goldstone equation in a uniform nuclear matter
 at given incident energy; the Bonn-B potential~\cite{Mac87} is adopted as a bare $NN$ interaction. In Ref.~\cite{Amos:2000}, the Melbourne $g$-matrix interaction was constructed and applied to a systematic investigation on proton elastic and inelastic scattering off various stable nuclei and some neutron-rich nuclei at energies from 40~MeV to 300~MeV. The nonlocality coming from the exchange term was rigorously treated and the central, spin-orbit, and tensor contributions were taken into account. As a result, it was clearly shown that the microscopic calculation with the Melbourne $g$ matrix for the proton-nucleus scattering has predictive power for the proton-nucleus elastic and inelastic cross sections and spin observables. Later, the framework was applied also to proton inelastic scattering off $^{10}$C and $^{18}$O~\cite{Karataglidis:2007yj}.

In the present study, we adopt a simplified single-folding model described in Ref.~\cite{Min10}. We employ the  Brieva and Rook (BR) prescription \cite{Brieva:1977rsh,Brieva:1977zz,Brieva:1978fdf}
to localize the exchange terms. The validity of the BR localization for nucleon-nucleus scattering was confirmed in Refs.~\cite{Min10,Hai16} and for nucleus-nucleus scattering in Ref.~\cite{Hag06}. This simplified single-folding model has successfully been applied to nucleon-nucleus elastic scattering for various cases~\cite{Min10,Toy13,Toyokawa:2015zxa,Minomo:2017hjl}.
In this study, we extend the model to proton inelastic scattering in a similar manner to in our recent studies on $\alpha$ inelastic scattering~\cite{Kanada-Enyo:2019prr,Kanada-Enyo:2019qbp}. To avoid complexity, we take into account only the central part of the proton-nucleus potential. The spin-orbit interaction is known to smear the dip structure of the diffraction pattern in general. Although at higher energies, say, above 150--200~MeV, it can somewhat affect the absolute amplitudes also near the peaks, such effect is expected to be minor in the energy region considered in this study. As in the previous studies including that by the Melbourne group~\cite{Amos:2000}, the local density approximation is adopted to apply the $g$-matrix interaction to a finite nucleus.

The cross sections of the elastic and inelastic scattering
are calculated by the CC calculations using the
proton-nucleus potentials obtained with the AMD densities for $^{18}$O, $^{10}$Be, $^{12}$Be, and $^{16}$C,
and the AMD+GCM densities for $^{12}$C and $^{16}$O. For $^{12}$C, we also use the
densities of a 3$\alpha$-cluster model of the resonating group method (RGM) \cite{Kamimura:1981oxj}.

It should be commented that a similar approach of the MCC calculation
with the Jeukenne-Lejeune-Mahaux (JLM) interaction \cite{Jeukenne:1977zz}
has been applied to the proton inelastic scattering off $^{10}$Be and $^{12}$Be in the earlier work by Takashina {\it et al}. \cite{Takashina:2008zza}. It was used also in continuum-discretized coupled-channels calculation for nucleon-induced breakup reactions of $^{6,7}$Li~\cite{Mat11,Ich12,Guo19} and $^{11}$Li~\cite{Mat17}.
The JLM interaction is another kind of the $g$-matrix effective interaction that has only the direct term. This property allows one to implement it to reaction calculation with the minimal task. On the other hand, in general, phenomenological tuning of the JLM interaction is necessary to fit the scattering data.

\subsection{Structure calculations for target nuclei}
\label{sec3c}
For the structure calculation of the target nuclei, we adopt the AMD wave functions
obtained by the variation after projections (VAP). In the AMD+VAP method, the variation is performed for the
spin-parity projected AMD wave function as done in Refs.~\cite{KanadaEn'yo:1998rf,Kanada-Enyo:2006rjf}.
The method was applied for the structure studies of $^{10}$Be, $^{12}$Be, and $^{16}$C
in Refs.~\cite{Kanada-Enyo:1999bsw,Kanada-Enyo:2003fhn,Kanada-Enyo:2004tao}. In the present paper,
the same method is
applied to $^{18}$O to obtain the wave functions of the $0^+_1$ and $2^+_1$ states.
For $^{12}$C and $^{16}$O, the AMD+VAP method is combined with the $3\alpha$- and
$^{12}\textrm{C}+\alpha$-cluster GCM, respectively,
as done in Refs.~\cite{Kanada-Enyo:2019prr,Kanada-Enyo:2019hrm,Kanada-Enyo:2019qbp,Kanada-Enyo:2015vwc,Kanada-Enyo:2017ers}.
In this paper, we simply call the AMD+VAP ``AMD'' and that with the cluster GCM ``AMD+GCM''.

The AMD wave functions used in this paper are in principle the same as those of Refs.~\cite{KanadaEn'yo:1998rf,Kanada-Enyo:2003fhn}.
We utilize the $^{10}$Be wave function for $^{10}$C by assuming the mirror symmetry.
For $^{16}$C, the VAP(c) wave function of Ref.~\cite{Kanada-Enyo:2004tao} is adopted.
The wave functions and transition densities of $^{12}$C and $^{16}$O are
consistent with those of AMD+GCM used for the $\alpha$ scattering
in Refs.~\cite{Kanada-Enyo:2019prr,Kanada-Enyo:2019qbp}.

The neutron and proton
matter and transition densities
are calculated with the AMD and AMD+GCM wave functions.
We denote the neutron and proton transition densities as $\rho^\textrm{tr}_{n}(r)$ and
$\rho^\textrm{tr}_{p}(r)$, respectively.
For $N=Z$ nuclei ($^{12}$C and $^{16}$O), half of isoscalar density is used as the
proton (neutron) density in the mirror symmetry assumption.
For quantitative discussions of inelastic cross sections, we scale the original
transition densities $\rho^\textrm{tr-cal}_{p}(r)$ to adjust the
theoretical $B(E\lambda)$ values to the experimental data as
\begin{align}
\rho^\textrm{tr}_{p}(r)=(M_{p}^\textrm{exp}/M^\textrm{cal}_{p})\rho^\textrm{tr-cal}_{p}(r)
\end{align}
Here the rank $\lambda$ ($\lambda >0 $) transition matrix elements for the neutron and proton parts
are defined as
\begin{align}
M_{n,p}\equiv \int r^{2+\lambda}\rho^\textrm{tr}_{n,p}(r) dr
\end{align}
and related to the transition strengths as
\begin{align}
B^{(n),(p)}_{\lambda} = \frac{1}{2J_i+1} |M_{n.p}|^2,
\end{align}
where $J_i$ is the angular momentum of the initial state.
The $E2$ transition strength is given by the proton $\lambda=2$ transition strength as 
$B(E2)=e^2B^{(p)}_{\lambda=2}$. 

The adopted states in the CC calculation for
$^{10}$Be, $^{12}$Be, $^{16}$C, and $^{18}$O are
$^{10}$Be$(0^+_{1,2},2^+_{1,2,3})$,  $^{12}$Be$(0^+_{1,2},2^+_{1,2})$,
$^{16}$C$(0^+_{1},2^+_{1,2})$, and $^{18}$O$(0^+_{1},2^+_{1})$.
All $\lambda=0$ and $\lambda=2$ transitions between these states are taken into account.
The experimental values of excitation energies are adopted as inputs of the
CC calculation.

In the CC calculations for $^{12}$C and $^{16}$O,
all the inputs from the structure part such as the adopted states, excitation energies, and transitions
are the same as those used for the $\alpha$ scattering  with the AMD+GCM wave functions
in Refs.~\cite{Kanada-Enyo:2019prr,Kanada-Enyo:2019qbp}.

As shown later, the CC effect gives only minor contribution
to the inelastic scattering to the $2^+_1$ state
at incident energies higher than 25 MeV,
and the cross sections are approximately described by
the one-step process of the distorted wave Born approximation (DWBA).

\section{Results of $^{12}$C and $^{16}$O}  \label{sec:results1}

\begin{figure}[!h]
\includegraphics[width=7 cm]{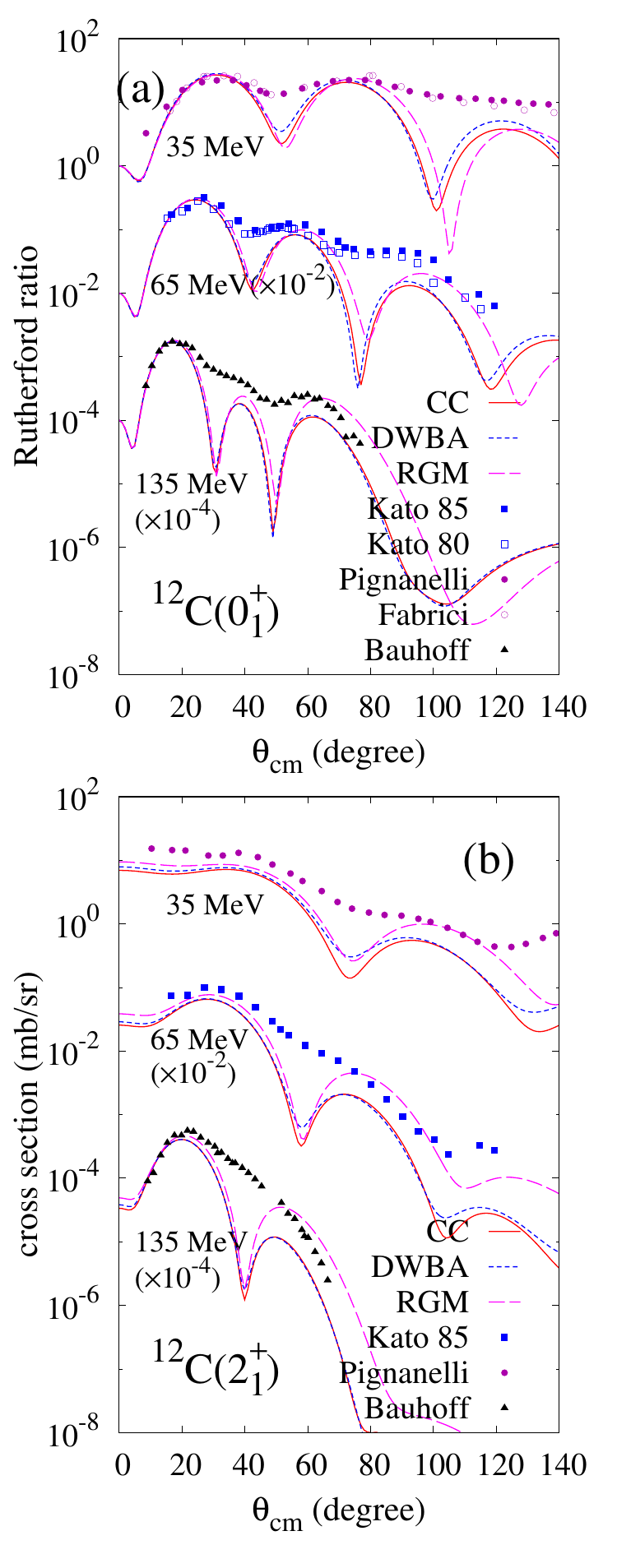}
\caption{
Cross sections of the
elastic and inelastic proton scattering off  $^{12}$C
at $E_p=35$ MeV, $E_p=65$ MeV $(\times 10^{-2})$, and $E_p=135$ MeV  $(\times 10^{-4})$
calculated with the AMG+GCM and RGM densities.
The results of the CC and DWBA calculations with the AMD+GCM densities and the CC calculation with the RGM densities
are shown by red solid, blue dotted, and magenta dashed lines, respectively.
The experimental data are from Refs.~\cite{Fabrici:1980zz,Pignanelli:1986zz,KATO1980589,Kato:1985zz,Bauhoff:1984kx}.
  \label{fig:cross-c12p}}
\end{figure}
\begin{figure}[!h]
\includegraphics[width=7 cm]{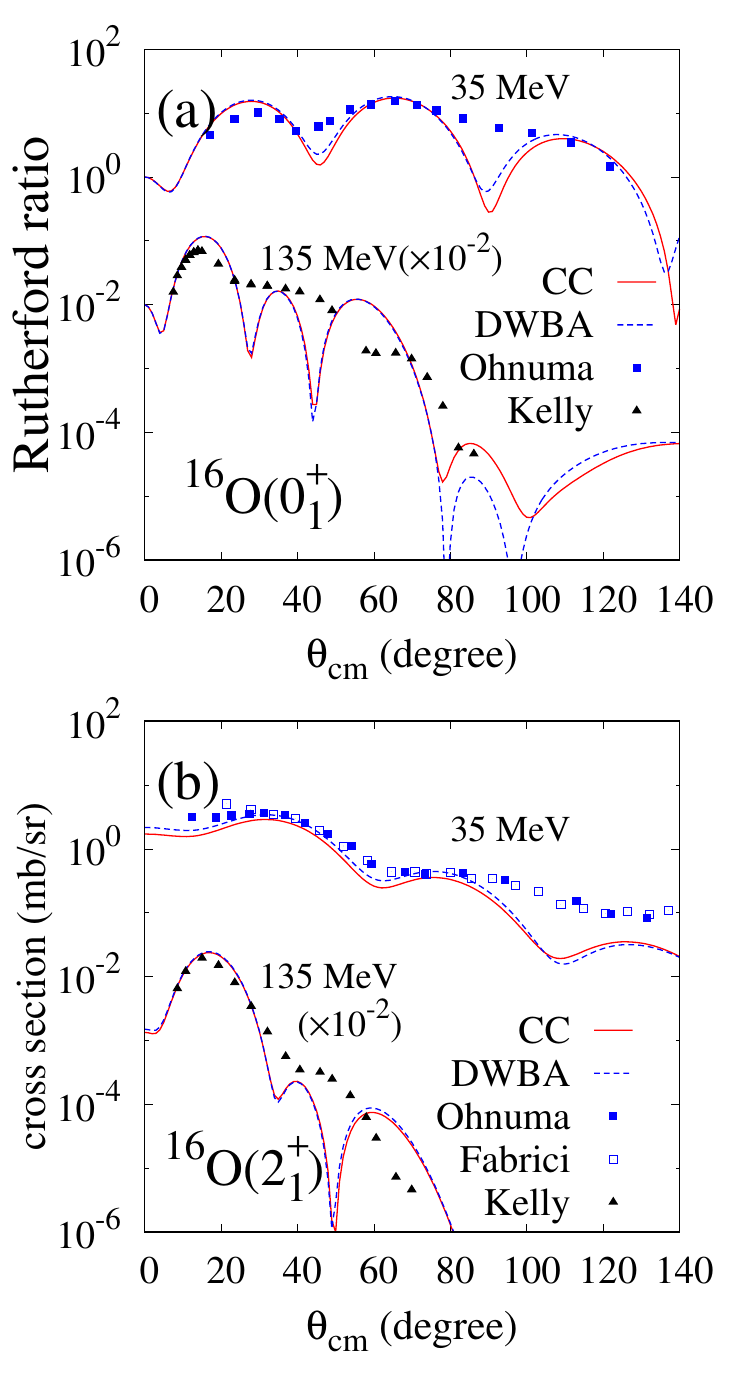}
  \caption{Cross sections of the
elastic and inelastic proton scattering off $^{16}$O
at $E_p=35$ MeV and $E_p=135$ MeV  $(\times 10^{-2})$
calculated with the AMG+GCM densities.
The results of the CC and  DWBA calculations are shown by red solid and blue dotted lines, respectively.
The experimental data are from Refs.~\cite{Fabrici:1980zz,Kelly:1980nd,Ohnuma:1990tkk}.
  \label{fig:cross-o16p}}
\end{figure}

The $0^+_1$ and $2^+_1$ cross sections of  $^{12}$C
at incident energies $E_p=35$, 65, and 135 MeV are shown in Fig.~\ref{fig:cross-c12p},
and those of $^{16}$O at $E_p=35$ and 135 MeV are shown in Fig.~\ref{fig:cross-o16p}.
In addition to the CC calculation, the one-step cross sections obtained by the DWBA calculation are also shown.
The small difference between the CC and DWBA cross sections indicates that the CC effect is minor.

The AMD+GCM result of $^{12}$C is compared with experimental data and also with the
calculation with the RGM density.
From the electron scattering data,
the RGM density is known to be good in quality and better than the AMD+GCM density
\cite{Kamimura:1981oxj,Kanada-Enyo:2019prr}.
As seen in Fig.~\ref{fig:cross-c12p},
the present calculation with the AMD+GCM density reproduces well the elastic proton scattering cross
sections of  $^{12}$C at forward angles, but somewhat underestimates the third peak.
A better result is obtained by the calculation with the RGM density, consistently with Ref.~\cite{Minomo:2017hjl}.
The inelastic proton scattering cross sections of $^{12}$C$(2^+_1)$ are described reasonably well with the
AMD+GCM and RGM calculations except for the cross sections at $E_p=35$ MeV.
The RGM density again gives a better agreement with the data at large angles.
This result indicates that quality of the structure model densities can be tested
by the detailed data of the proton scattering.
For the proton scattering off $^{16}$O,  the present calculation reproduces well the elastic and inelastic  cross sections
 (Fig.~\ref{fig:cross-o16p}).
It should be commented that
the $2^+_1$ state of $^{16}$O is not the ground-band member but belongs to the
$^{12}$C+$\alpha$-cluster band built on the $0^+_2$ state.
The present microscopic approach works well
even for such the developed cluster state with structure much different from the ground state.

\section{Results of $Z\ne N$ nuclei}  \label{sec:results2}

\subsection{Structure properties}

The theoretical  and experimental values of structure properties of the target nuclei are listed in Tables~\ref{tab:energy}
and \ref{tab:radii}.
The energies are shown in Table~\ref{tab:energy}, and radii and $\lambda=2$ transition strengths
as well as the $M_n/M_p$ ratio
are shown in
Table~\ref{tab:radii}.
The structure calculation of $^{10}$Be, $^{12}$Be, and $^{16}$C are consistent with
Refs.~\cite{Kanada-Enyo:1999bsw,Kanada-Enyo:2003fhn,Kanada-Enyo:2004tao}.
We will describe detailed properties of the $0^+_1$ and $2^+_1$ states and the
transition between them later.

\begin{table}[ht]
\caption{Binding and excitation energies of $^{18}$O, $^{10}$Be, $^{12}$Be, and $^{16}$C.
Theoretical values of $^{10}$Be, $^{12}$Be, and $^{16}$C are taken from
Refs.~\cite{Kanada-Enyo:1999bsw,Kanada-Enyo:2003fhn,Kanada-Enyo:2004tao}, and experimental values
are from Refs.~\cite{Tilley:2004zz,Kelley:2017qgh,Tilley:1993zz,Tilley:1995zz}.
The band assignment ($K^\pi$) are given based on the AMD calculation.
 \label{tab:energy}
}
\begin{center}
\begin{tabular}{lcccccccc}
\hline
 &   \multicolumn{2}{c}{Energy (MeV) }\\
 &Band &  AMD & exp\\
$^{18}$O$(0^+_1)$   & $K=0^+_1$ & 131.1   & 139.80  \\
$^{18}$O$(2^+_1)$   & $K=0^+_1$ & 2.0   & 1.98  \\
  &   &   &   \\
$^{10}$Be$(0^+_1)$  & $K=0^+_1$ & 61.1  & 64.98 \\
$^{10}$Be$(2^+_1)$  & $K=0^+_1$ & 2.7   & 3.37  \\
$^{10}$Be$(2^+_2)$  & $K=2^+$ & 6.8   & 5.96  \\
$^{10}$Be$(0^+_2)$  & $K=0^+_2$ & 7.8   & 6.179 \\
$^{10}$Be$(2^+_3)$  & $K=0^+_2$ & 9.0   & 7.54  \\
  &   &   &   \\
$^{12}$Be$(0^+_1)$  & $K=0^+_1$ & 61.9  & 68.65 \\
$^{12}$Be$(2^+_1)$  & $K=0^+_1$ & 1.8   & 2.11  \\
$^{12}$Be$(0^+_2)$  & $K=0^+_2$ & 3.6   & 2.251 \\
$^{12}$Be$(2^+_2)$  & $K=0^+_2$ & 4.6   & $-$ \\
  &   &   &   \\
$^{16}$C$(0^+_1)$ & $K=0^+_1$ & 102.6   & 110.75  \\
$^{16}$C$(2^+_1)$ & $K=0^+_1$ & 2.4   & 1.77  \\
$^{16}$C$(2^+_2)$ & $K=2^+$ & 7.8   & 3.99  \\
\hline
\end{tabular}
\end{center}
\end{table}

\begin{table}[ht]
\caption{Matter, proton, and neutron
radii, and transition strengths of  $^{18}$O, $^{10}$Be, $^{12}$Be, and $^{16}$C.
Theoretical values of the AMD calculation for $^{10}$Be, $^{12}$Be, and $^{16}$C are from
Refs.~\cite{Kanada-Enyo:1999bsw,Kanada-Enyo:2003fhn,Kanada-Enyo:2004tao}, and experimental values
are taken from Refs.~\cite{Tilley:2004zz,Kelley:2017qgh,Tilley:1993zz,Tilley:1995zz}.
The data of $B^{(p)}_{\lambda=2}=B(E2)/e^2$ for $^{16}$C are the values reported in Refs.~\cite{Ong:2007jb, Wiedeking:2008zzb}.
$^*$  The experimental values of $B(E2)/e^2$ of the mirror nuclei ($^{18}$Ne and $^{10}$C) are shown
for $B^{(n)}_{\lambda=2}$ of $^{18}$O and $^{10}$Be.
 \label{tab:radii}
}
\begin{center}
\begin{tabular}{llllllllcccccccc}
\hline
  & $R_p$ (fm)  & $R_n$ (fm)  & $R_m$ (fm)  \\
$^{18}$O$(0^+_1)$   & 2.75  & 2.88  & 2.82  \\
exp & 2.62  & 2.83$^\textrm{*mir}$  & 2.61(8) \\
  & $B^{(p)}_{\lambda=2}$ (fm$^4$)  & $B^{(n)}_{\lambda=2}$ (fm$^4$)  & $M_n/M_p$ \\
$^{18}$O$(2^+_1 \to 0^+_1)$   & 0.7   & 18.6  & 5.4   \\
exp & 9.3(3)  & 50(5)$^\textrm{*mir}$ & 2.3(2)$^\textrm{*mir}$  \\
\hline
  & $R_p$ (fm)  & $R_n$ (fm)  & $R_m$ (fm)  \\
$^{10}$Be$(0^+_1)$  & 2.50  & 2.56  & 2.54  \\
exp & 2.17  &   & 2.30(2) \\
  & $B^{(p)}_{\lambda=2}$ (fm$^4$)  & $B^{(n)}_{\lambda=2}$ (fm$^4$)  & $M_n/M_p$ \\
$^{10}$Be$(2^+_1 \to 0^+_1)$  & 11.6  & 8.9   & 0.9   \\
exp & 10.2(1.0) & 12.2(1.9)$^\textrm{*mir}$ & 1.1(1)$^\textrm{*mir}$  \\
$^{10}$Be$(2^+_2 \to 0^+_1)$  & 0.2   & 3.2   & 3.9   \\
$^{10}$Be$(2^+_3 \to 0^+_1)$  & 0.1   & 0.7   & 2.5   \\
$^{10}$Be$(2^+_3 \to 0^+_2)$  & 34.5  & 118   & 1.8   \\
\hline
  & $R_p$ (fm)  & $R_n$ (fm)  & $R_m$ (fm)  \\
$^{12}$Be$(0^+_1)$  & 2.67  & 2.94  & 2.85  \\
exp & 2.39  &   & 2.59(6) \\
$^{12}$Be$(0^+_2)$  & 2.56  & 2.84  & 2.75  \\
  & $B^{(p)}_{\lambda=2}$ (fm$^4$)  & $B^{(n)}_{\lambda=2}$ (fm$^4$)  & $M_n/M_p$ \\
$^{12}$Be$(2^+_1 \to 0^+_1)$  & 14.4  & 51.1  & 1.9   \\
exp & 14.2(2.8) &   &   \\
$^{12}$Be$(2^+_2 \to 0^+_1)$  & 0.0   & 7.4   & 25.4  \\
$^{12}$Be$(2^+_2 \to 0^+_2)$  & 7.5   & 9.0   & 1.1   \\
\hline
  & $R_p$ (fm)  & $R_n$ (fm)  & $R_m$ (fm)  \\
$^{16}$C$(0^+_1)$   & 2.58  & 2.85  & 2.75  \\
exp &   &   & 2.70(3) \\
  & $B^{(p)}_{\lambda=2}$ (fm$^4$)  & $B^{(n)}_{\lambda=2}$ (fm$^4$)  & $M_n/M_p$ \\
$^{16}$C$(2^+_1 \to 0^+_1)$   & 2.7   & 27.0  & 3.2   \\
exp \cite{Ong:2007jb} & 2.6(9)  &   &   \\
exp \cite{Wiedeking:2008zzb}  & 4.15(73)  &   &   \\
$^{16}$C$(2^+_2 \to 0^+_1)$   & 2.6   & 0.1   & 0.2   \\
\hline
\end{tabular}
\end{center}
\end{table}

\subsection{Results of $^{18}$O}

\begin{figure}[!h]
\includegraphics[width=7 cm]{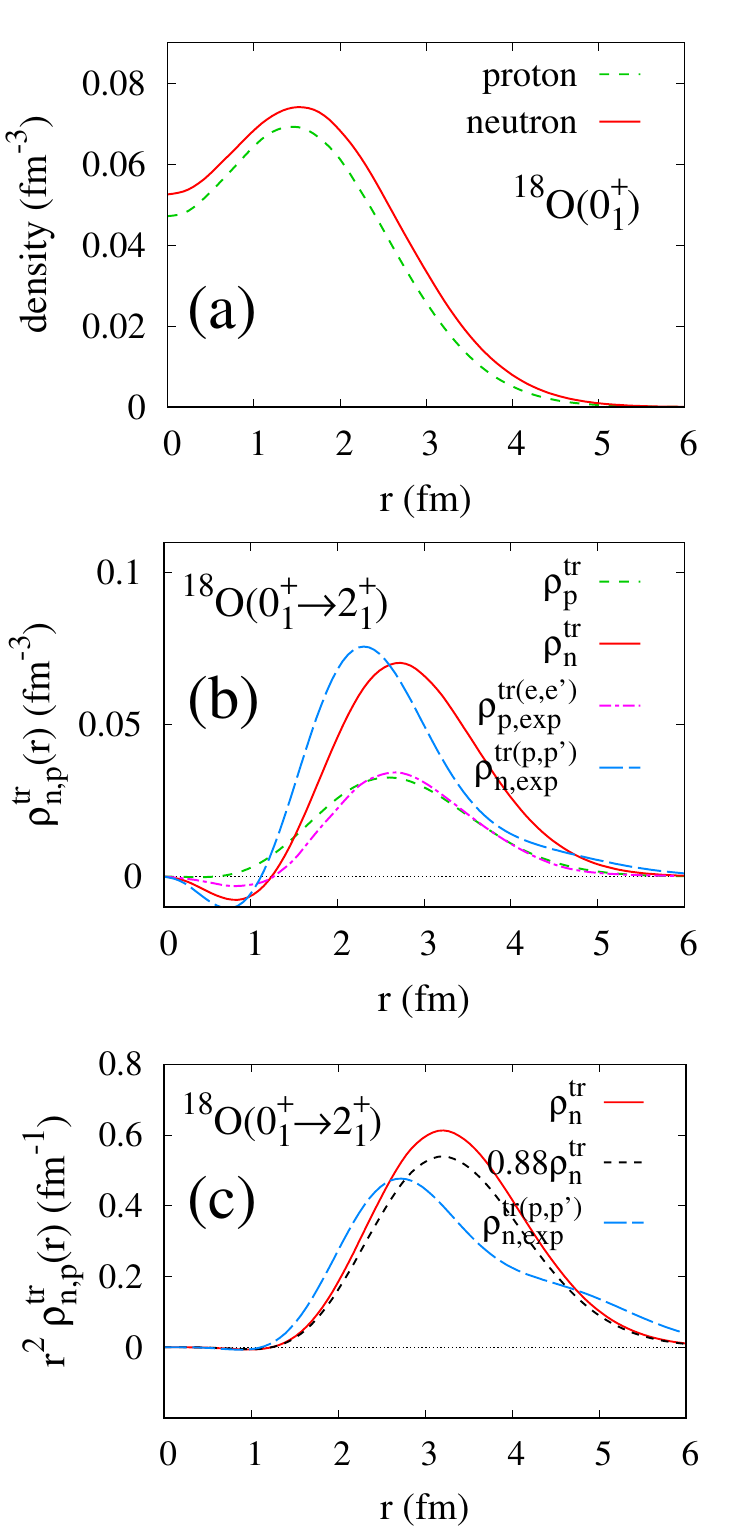}
  \caption{Neutron and proton densities of $^{18}\textrm{O}$.
(a) The neutron and proton matter densities of the ground state, (b)
the neutron and proton transition densities for the $0^+_1\to 2^+_1$ transition,
and (c) the $r^2$-weighted neutron transition density calculated with AMD.
The renormalized proton and neutron transition densities adjusted to the experimental $B(E2)$ of
$^{18}\textrm{O}$ and that of $^{18}\textrm{Ne}$ are shown, respectively.
The experimental neutron transition density $\rho^{\textrm{tr}(p,p')}_{n,\textrm{exp}}$  reduced
from the $(p,p')$ scattering at $E=135$ MeV/u \cite{Kelly:1986ysn} and the experimental proton transition density
$\rho^{\textrm{tr}(e,e')}_{p,\textrm{exp}}$ measured with the electron scattering data \cite{Norum:1982cj}
are also shown.
  \label{fig:trans-o18}}
\end{figure}

\begin{figure}[!h]
\includegraphics[width=6 cm]{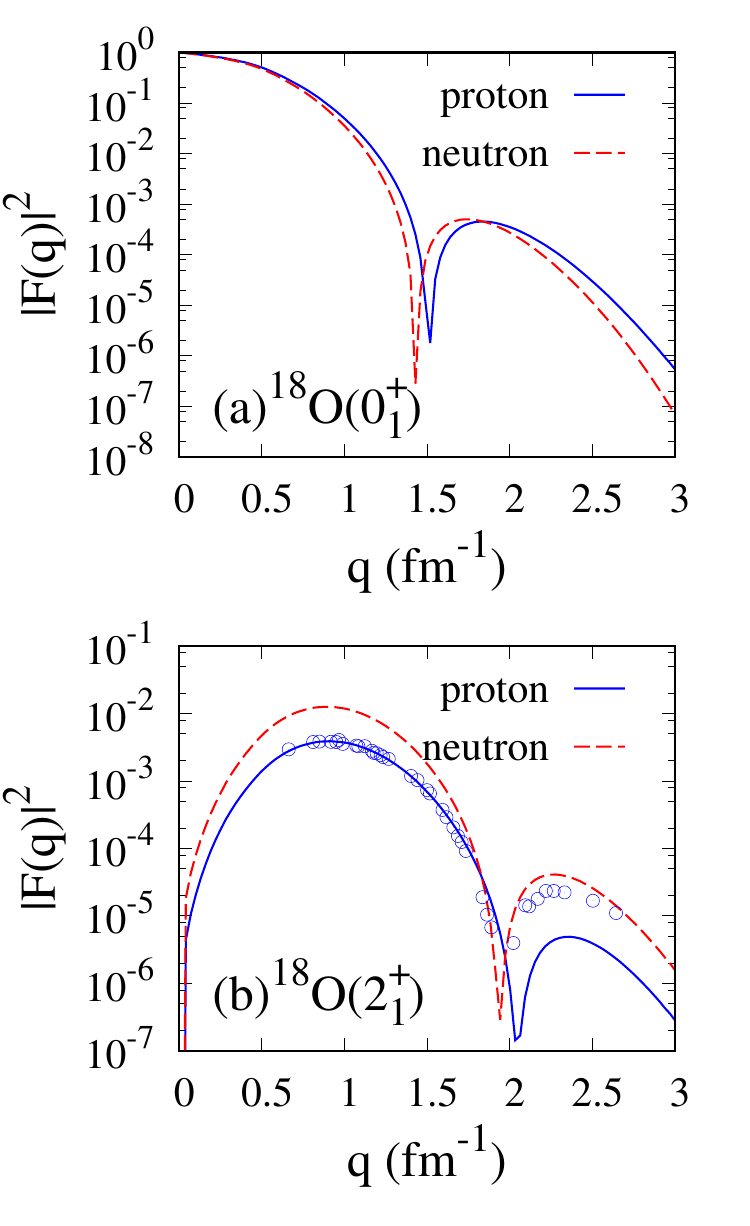}
  \caption{Elastic and inelastic form factors of $^{18}\textrm{O}$.
The inelastic form factors for $0^+_1\to 2^+$ are the renormalized ones adjusted to
experimental $B(E\lambda)$.  The experimental data measured by the electron scattering are from
Ref.~\cite{Norum:1982cj}.
  \label{fig:form-o18}}
\end{figure}

\begin{figure}[!h]
\includegraphics[width=7 cm]{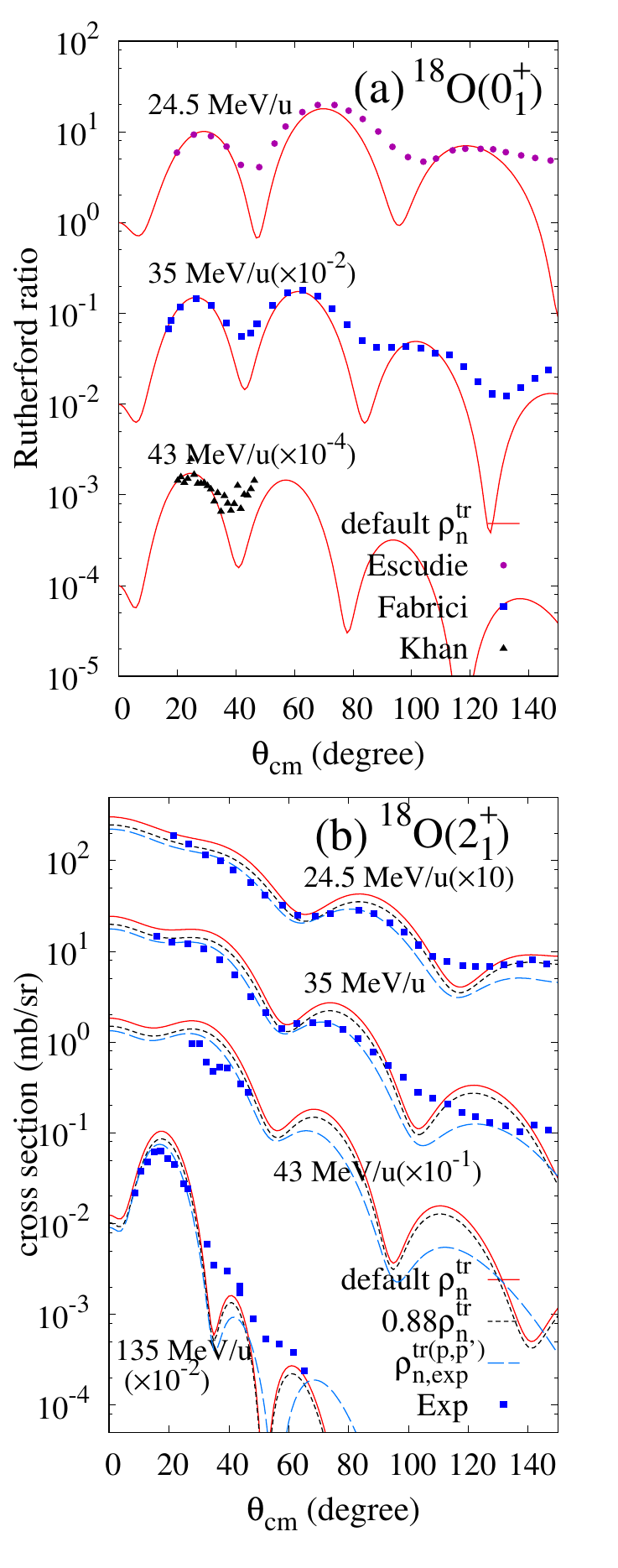}
\caption{
Cross sections of the elastic and inelastic proton scattering off $^{18}$O
at $E=24$ MeV/u ($\times 10$), 35  MeV/u, 43 MeV/u ($\times 10^{-1}$), and 135 MeV/u $(\times 10^{-2})$
calculated with the default AMD densities (solid lines).
The experimental data for $E=43$ MeV/u are the cross sections measured
in inverse kinematics.
For the $2^+_1$ cross sections, the calculated result with the
experimental neutron transition density $\rho_{n,\textrm{exp}}^{\textrm{tr}(p,p')}(r)$
and that with the reduced neutron transition density $0.88\rho^\textrm{tr}_n(r)$ are also shown by dashed and dotted lines, respectively.
The experimental data are from
Refs.~\cite{Escudie:1974zz,Kelly:1986ysn,Fabrici:1980zz,Khan:2000rac}
  \label{fig:cross-o18p}}
\end{figure}

\begin{figure}[!h]
\includegraphics[width=7 cm]{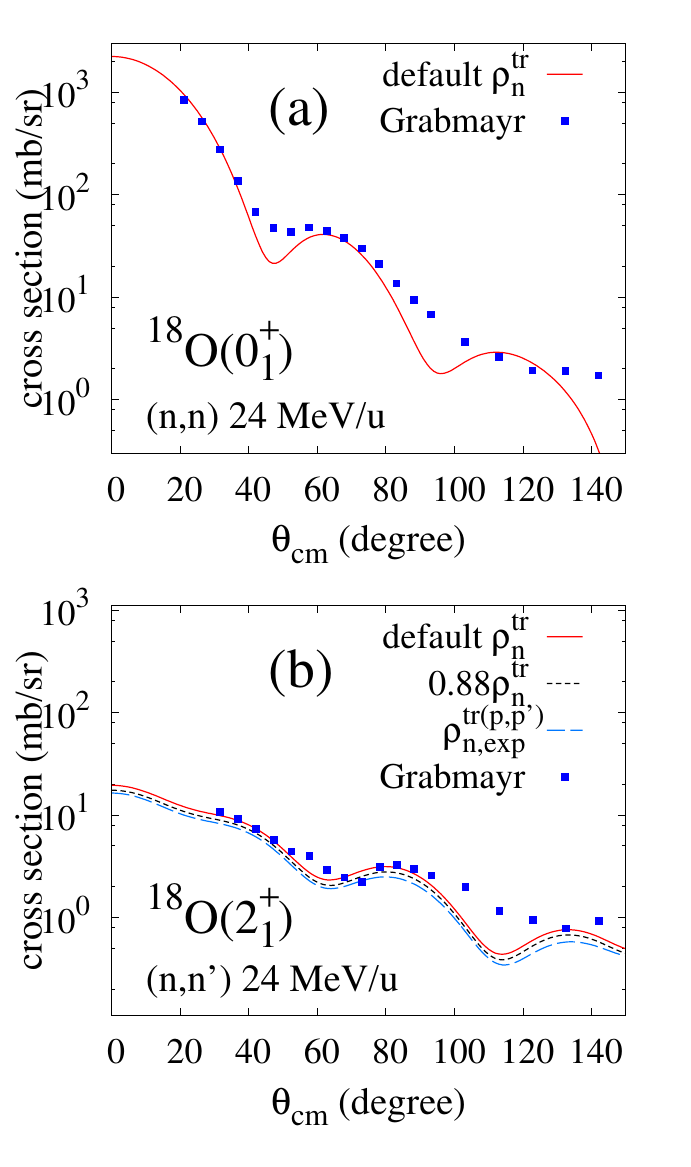}
  \caption{Cross sections of (a) elastic and (b) inelastic neutron scattering off $^{18}\textrm{O}$
at $E=24$ MeV/u.
For the $2^+_1$ cross sections, the CC calculation with the
experimental neutron transition density $\rho_{n,\textrm{exp}}^{\textrm{tr}(p,p')}(r)$
and that with $0.88\rho^\textrm{tr}_n(r)$ are also shown by blue dotted and light-blue dashed lines, respectively, 
in addition to that with the default AMD densities (red solid lines). 
The data are from Ref.~\cite{Grabmayr:1980qze}.
  \label{fig:cross-o18n}}
\end{figure}

\begin{figure}[!h]
\includegraphics[width=7 cm]{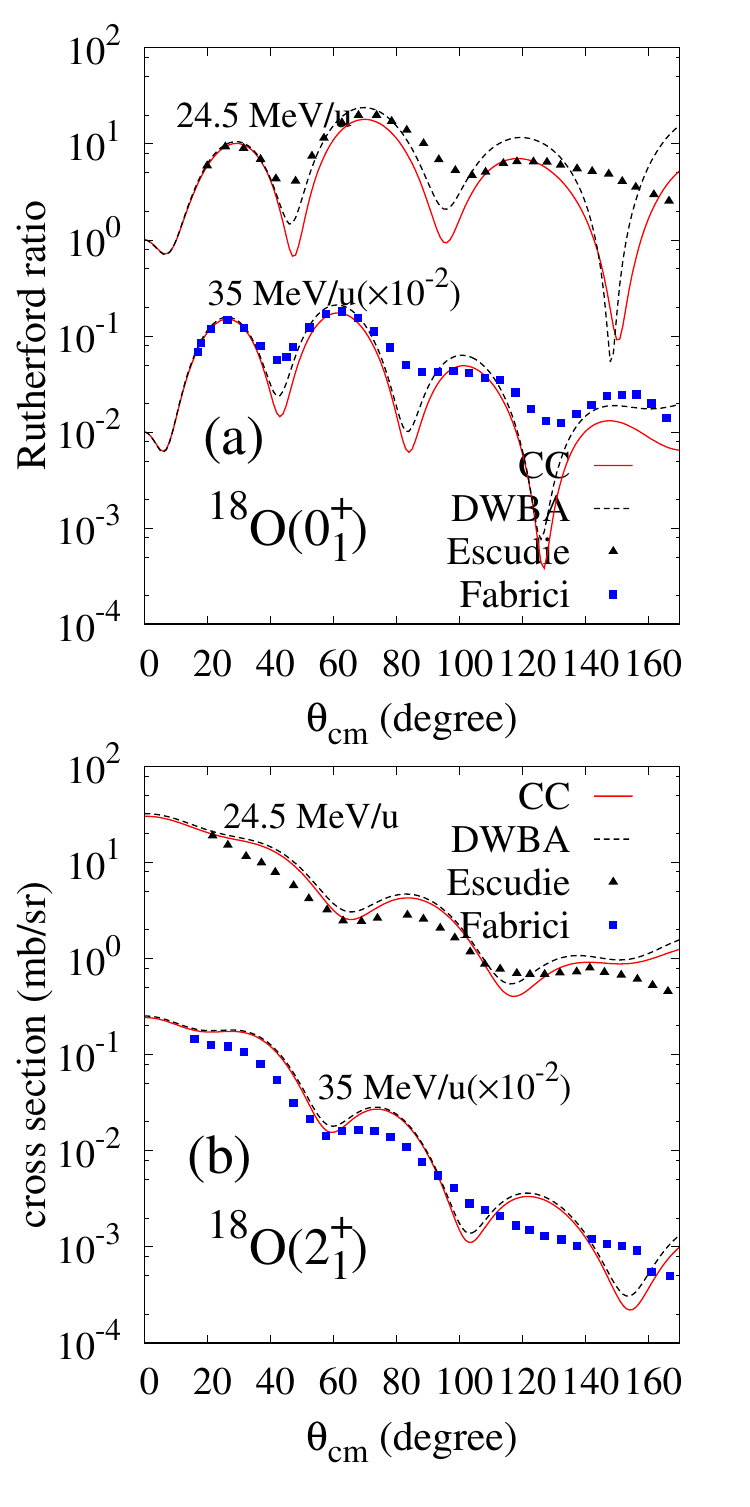}
\caption{Comparison of the CC and DWBA calculations
of the proton scattering off $^{18}$O with the default $\rho^\textrm{tr}_n(r)$.
The (a) elastic and (b) inelastic cross sections at $E=24.5$ MeV/u and 35  MeV/u are shown
in comparison with the experimental data \cite{Escudie:1974zz,Fabrici:1980zz}.
  \label{fig:cross-o18p-dwba}}
\end{figure}

In the $2^+_1\to 0^+_1$ transition of  $^{18}$O, the significant $B^{(p)}_{\lambda=2}$ has been experimentally
known but it is much underestimated by the AMD calculation meaning that
the proton excitation from the $p$-shell closure is not sufficiently described in the theory.
For the neutron part, the large $B^{(n)}_{\lambda=2}$ of the AMD calculation indicates
the neutron dominance, which is qualitatively
consistent with the mirror transition of  $^{18}$Ne \cite{Bernstein:1979zza}
and the proton scattering experiment \cite{Kelly:1986ysn}.

The calculated densities and form factors of $^{18}$O are shown in Figs.~\ref{fig:trans-o18} and \ref{fig:form-o18},
respectively, together with the data measured by the electron scattering experiments.
Here, the theoretical proton transition density
$\rho^\textrm{tr-cal}_p(r)$ and
form factors $F^\textrm{cal}(q)$ are scaled by the factor
$M^\textrm{exp}_p/M^\textrm{cal}_p=3.88$ as
$\rho^\textrm{tr}_p(r)=(M^\textrm{exp}_p/M^\textrm{cal}_p)\rho^\textrm{tr-cal}_p(r)$ and
$F(q)=(M^\textrm{exp}_p/M^\textrm{cal}_p)F^\textrm{cal}(q)$
so as to fit the experimental $B(E2)$ value.
After the scaling, the experimental data are reproduced well
except for the small $r$ (large $q$) region.

For the neutron transition density $\rho^\textrm{tr}_n(r)$ of $^{18}\textrm{O}$, we tentatively
assume the mirror symmetry and
scale $\rho^\textrm{tr-cal}_n(r)$
with the scaling factor $M^\textrm{exp}_p(^{18}\textrm{Ne})/M^\textrm{cal}_n=1.72$
to prepare the default input $\rho^\textrm{tr}_n(r)$ in the reaction calculation. 
However, if we take into account
 the mirror symmetry breaking,
another choice maybe possible, for example, about a 10\% smaller value than $M^\textrm{exp}_p(^{18}\textrm{Ne})$
was theoretically recommended for $M_n(^{18}\textrm{O})$ in Ref.~\cite{Kelly:1986ysn}.
In Figs.~\ref{fig:trans-o18}(b) and (c),
the default $\rho^\textrm{tr}_n(r)$ is compared with the
experimental estimation (denoted by $\rho_{n,\textrm{exp}}^{\textrm{tr}(p,p')}(r)$) of Ref.~\cite{Kelly:1986ysn},
which  was reduced from the inelastic proton scattering at $E=135$ MeV/u by a model analysis.
$\rho_{n,\textrm{exp}}^{\textrm{tr}(p,p')}(r)$  gives  $B^{(n)}_{\lambda=2}=38$ fm$^4$, which is slightly smaller than
$B^{(n)}_{\lambda=2}=50$ fm$^4$ of $\rho^\textrm{tr}_n(r)$ adjusted to $B(E2;^{18}\textrm{Ne})$.
The 12\% reduced transition density (0.88$\rho^\textrm{tr}_n(r)$) gives the same strengths ($B^{(n)}_{\lambda=2}=38$ fm$^4$)
as $\rho_{n,\textrm{exp}}^{\textrm{tr}(p,p')}(r)$,
but it shows a different radial behavior from $\rho_{n,\textrm{exp}}^{\textrm{tr}(p,p')}(r)$.
Compared with the theoretical transition density,
$\rho^{\textrm{tr}(p,p')}_{n,\textrm{exp}}(r)$ has the smaller amplitude
at the nuclear surface ($r=3$--$4$ fm) and enhanced outer tail in $r \gtrsim 5$ fm region (see Fig.~\ref{fig:trans-o18}(c)).
In the reaction analysis,
we consider this difference between $\rho_{n,\textrm{exp}}^{\textrm{tr}(p,p')}(r)$  and the default $\rho^\textrm{tr}_n(r)$
as a model ambiguity from the neutron transition density.

We calculate the cross sections of the
proton scattering at $E=24.5$ MeV/u, $35$ MeV/u, 43 MeV/u, and 135 MeV/u, and those of the neutron scattering at
$E=24$ MeV/u. They are compared with the experimental data.
The results are shown in Figs.~\ref{fig:cross-o18p} and \ref{fig:cross-o18n}.
The calculation reproduces reasonably well
the elastic and inelastic proton scattering cross sections in the wide range of $E=24$--$135$ MeV/u.
It also reproduces well the neutron scattering cross sections at $E=24$ MeV/u.
In comparison with the DWBA calculation shown in Fig.~\ref{fig:cross-o18p-dwba},
one can see that the CC effect is minor in the $2^+_1$ cross sections.

Let us discuss the ambiguity from the proton and neutron transition densities.
As shown previously, the (scaled) proton part $\rho^\textrm{tr}_p(r)$ used in the present calculation reproduces
well the experimental data measured by the electron scattering, whereas
the neutron part $\rho^\textrm{tr}_n(r)$ has the $r$ behavior different from
the experimental one $\rho_{n,\textrm{exp}}^{\textrm{tr}(p,p')}(r)$.
In order to see the effect of this difference in the neutron transition density to
the inelastic  cross sections, we perform the same reaction calculation using
$\rho_{n,\textrm{exp}}^{\textrm{tr}(p,p')}(r)$  and $0.88\rho^\textrm{tr}_n(r)$.
Figures \ref{fig:cross-o18p}(b) and \ref{fig:cross-o18n}(b) show respectively
the proton and neutron 
scattering cross sections obtained with $\rho_{n,\textrm{exp}}^{\textrm{tr}(p,p')}(r)$
(light blue dashed lines) 
and that with $0.88\rho^\textrm{tr}_n(r)$ 
(blue dotted lines) 
in comparison with the default calculation 
(red solid lines) and experimental data.
In the result of the proton scattering with $\rho_{n,\textrm{exp}}^{\textrm{tr}(p,p')}(r)$, the cross sections at forward angle slightly decrease
to 70\% of the default calculation, and the second and third peaks at the large angles are reduced further to $40$--$60\%$
of the default calculation. 
The reduction rate at large angels is larger than the naive expectation of $38/50\approx 75\%$
from the $B^{(n)}_{\lambda=2}$ ratio.
It means that the outer tail amplitude of the neutron transition density gives relatively minor contribution to the proton scattering 
cross sections than the surface amplitude
though it significantly enhances the $M_n$, i.e., $B^{(n)}_{\lambda=2}$.
The calculation with $\rho_{n,\textrm{exp}}^{\textrm{tr}(p,p')}(r)$  obtains better agreement with the proton scattering 
data at least at $E=24.5$ MeV/u and 35 MeV/u suggesting that 
$\rho_{n,\textrm{exp}}^{\textrm{tr}(p,p')}(r)$  may be favored rather than the default
$\rho^\textrm{tr}_n(r)$ used in the present calculation.
It indicates that 
the proton scattering is a sensitive probe for the neutron transition density.
In contrast to the  proton scattering, the neutron scattering
cross sections are not so sensitive to 
the difference in the neutron transition densities 
as expected from the weaker $nn$ interactions than the 
$pn$ ones.

\subsection{Results of $^{10}$Be, $^{12}$Be,  and $^{16}$C}

\begin{figure}[!h]
\includegraphics[width=7 cm]{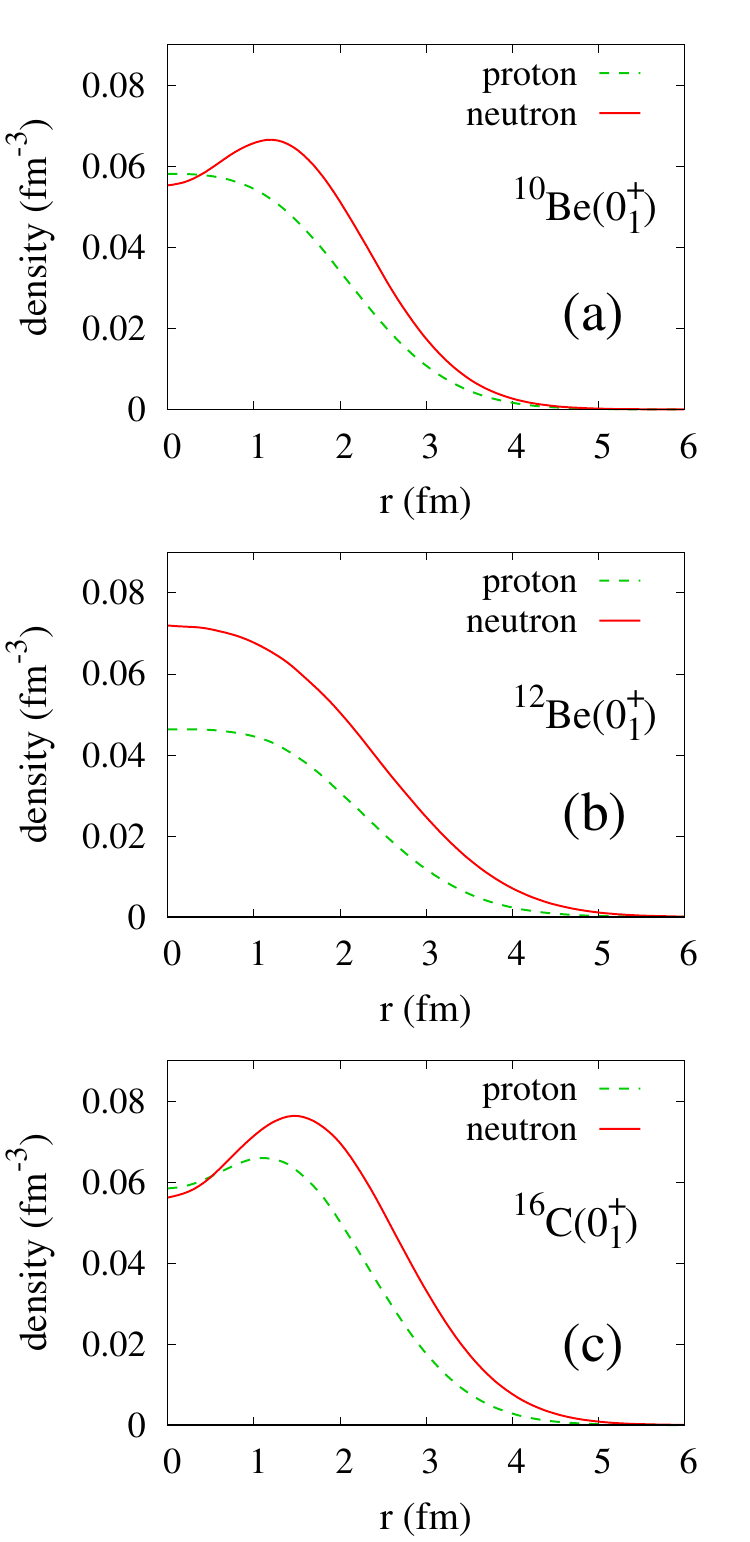}
  \caption{Neutron and proton matter densities of the ground states of (a) $^{10}\textrm{Be}$, (b) $^{12}\textrm{Be}$,
and (c) $^{16}\textrm{C}$ calculated with AMD.
  \label{fig:dens-be12}}
\end{figure}
\begin{figure}[!h]
\includegraphics[width=7 cm]{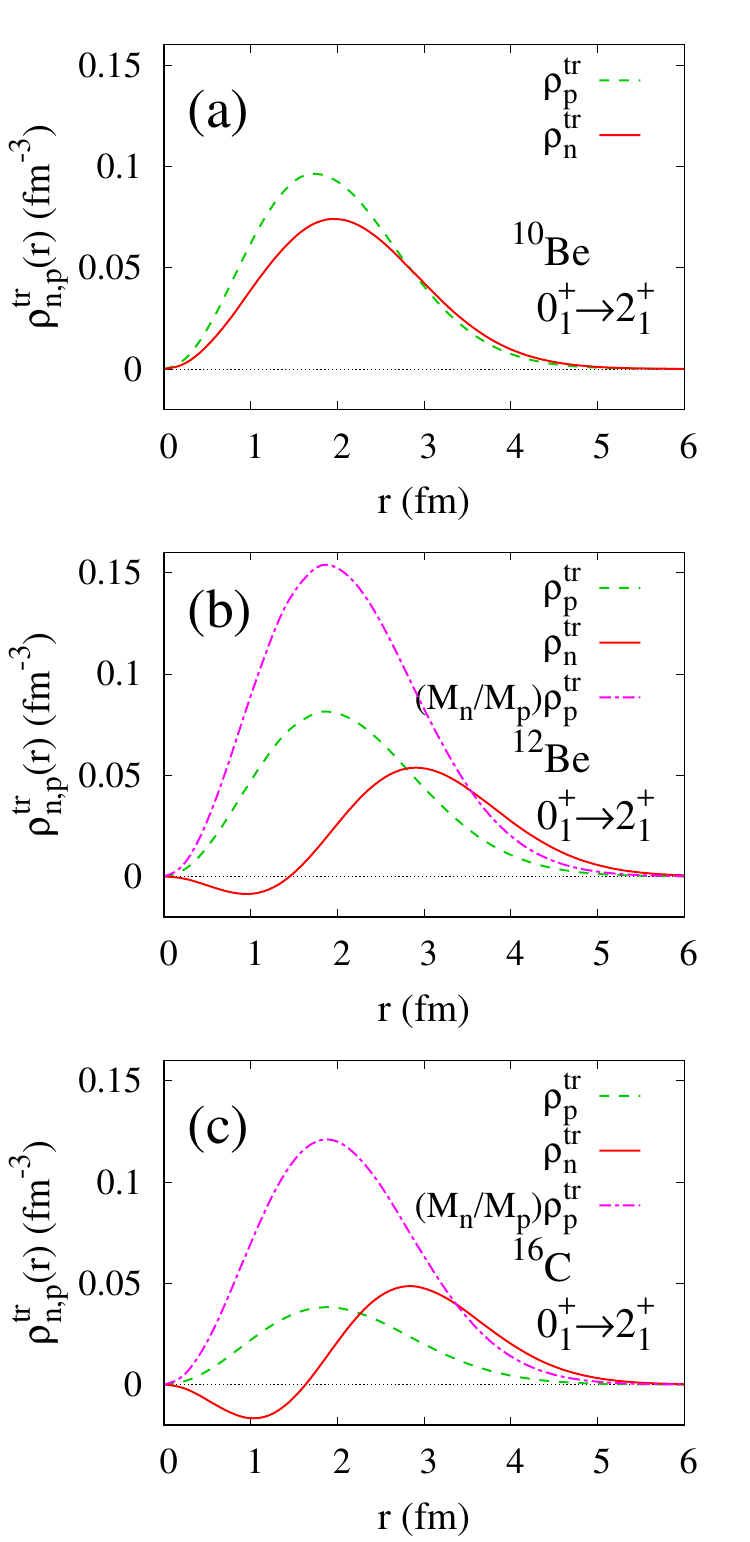}
  \caption{Neutron ($\rho^\textrm{tr}_n$) and proton ($\rho^\textrm{tr}_p$) transition densities for $0^+_1\to 2^+_1$ of
(a) $^{10}\textrm{Be}$, (b) $^{12}\textrm{Be}$, and (c) $^{16}\textrm{C}$ calculated with AMD.
The proton and neutron transition densities of $^{10}$Be
are renormalized to adjust the experimental $B(E2)$ value
of $^{10}\textrm{Be}$ and that of $^{10}\textrm{C}$, respectively.
\label{fig:trans-be12}}
\end{figure}
\begin{figure}[!h]
\includegraphics[width=7 cm]{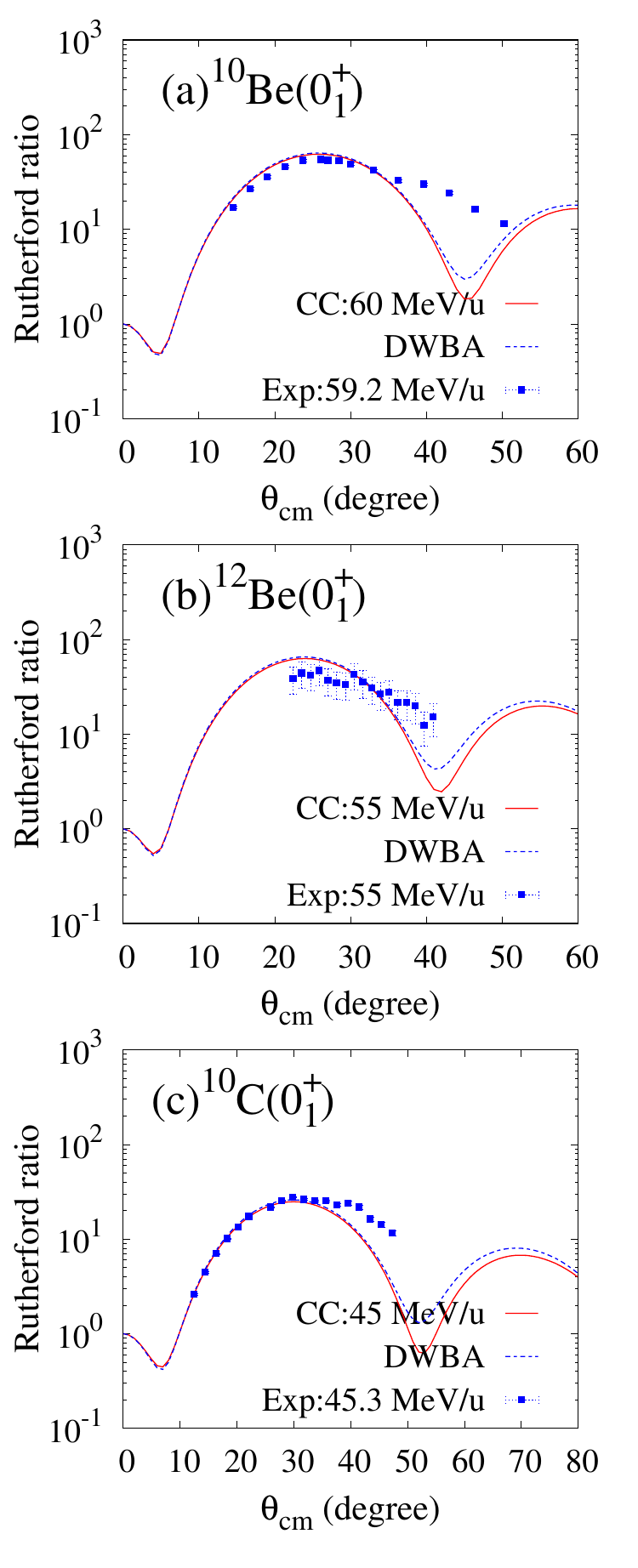}
  \caption{
Cross sections of the
elastic proton scattering off  (a) $^{10}\textrm{Be}$ at $E=60$ MeV/u, (b) $^{12}\textrm{Be}$ at $E=55$ MeV/u,
and (c) $^{10}\textrm{C}$ at $E=45$ MeV/u
calculated by the CC calculation with the AMD densities  (red solid lines).
The one-step cross sections obtained by the DWBA calculation are also shown (blue dotted lines).
The calculations are compared with the experimental data measured in inverse kinematics
of $^{10}\textrm{Be}$ at 59.2 MeV/u\cite{CortinaGil:1997zk},
$^{12}\textrm{Be}$ at 55 MeV/u\cite{Korsheninnikov:1995jtx},
and  $^{10}\textrm{C}$ at 45.3 MeV/u\cite{Jouanne:2005pb}.
\label{fig:cross-be12-ela}}
\end{figure}

\begin{figure}[!h]
\includegraphics[width=7 cm]{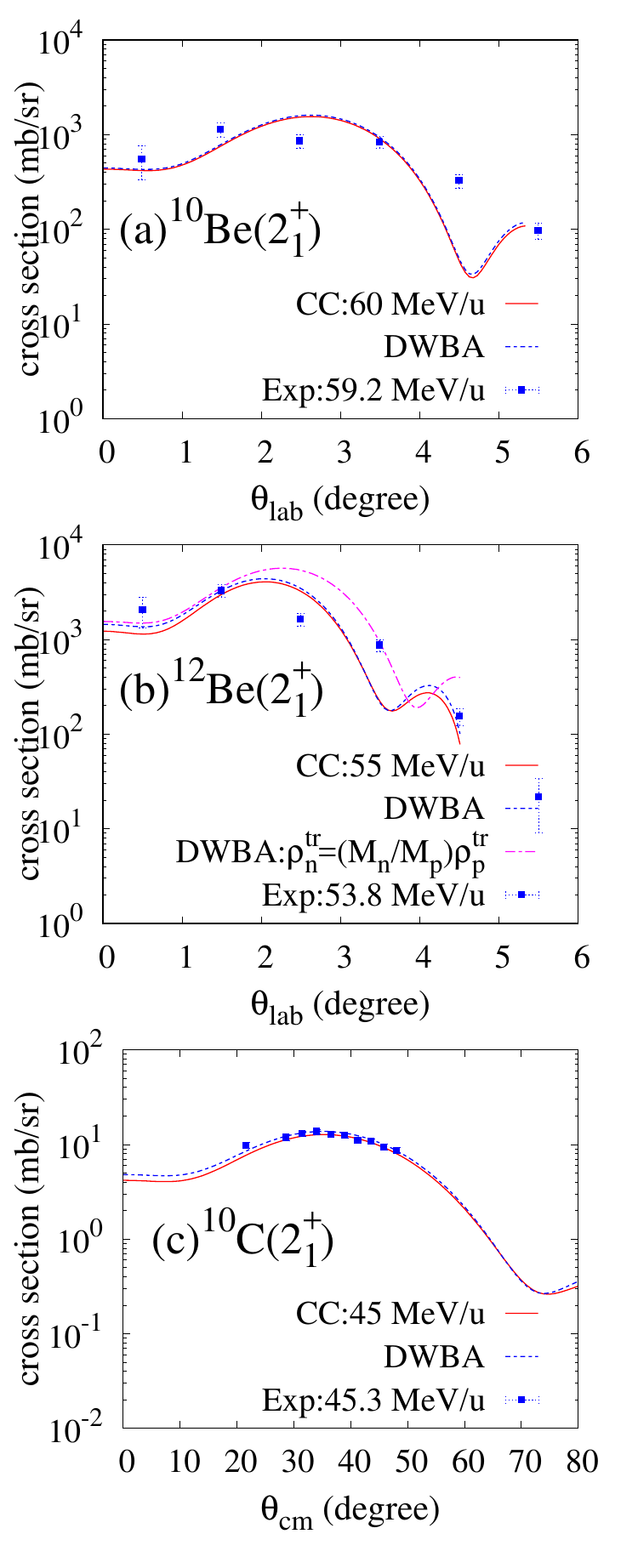}
  \caption{Cross sections of the
inelastic proton scattering to the $2^+_1$ state of (a) $^{10}\textrm{Be}$ at $E=60$ MeV/u, (b) $^{12}\textrm{Be}$ at $E=55$ MeV/u,
and (c) $^{10}\textrm{C}$ at $E=45$ MeV/u
calculated by the CC calculation (red solid lines).
The one-step cross sections obtained by the DWBA calculation are also shown (blue dotted lines).
In the panel (b) for $^{12}\textrm{Be}$, 
the DWBA calculation using the neutron transition density $\rho^\textrm{tr}_n(r)=(M_n/M_p) \rho^\textrm{tr}_p(r)$
is also shown for comparison (a magenta dash-dotted line).
The calculations are compared with the experimental data measured in inverse kinematics
of $^{10}\textrm{Be}$ at 59.2 MeV/u\cite{Iwasaki:2000gh},
$^{12}\textrm{Be}$ at 53.8 MeV/u\cite{Iwasaki:2000gh},
and  $^{10}\textrm{C}$ at 45.3 MeV/u\cite{Jouanne:2005pb}.
For the inelastic scattering of $^{10}\textrm{Be}$($^{12}\textrm{Be}$),
$\theta_\textrm{lab}$
is kinematically limited within 5.6 (4.7) degrees,
but the data
contain effects of finite size and angular spread of the incident beam, multiple scattering in the target, 
and detector geometry.
\label{fig:cross-be12-inel}}
\end{figure}

\begin{figure}[!h]
\includegraphics[width=7 cm]{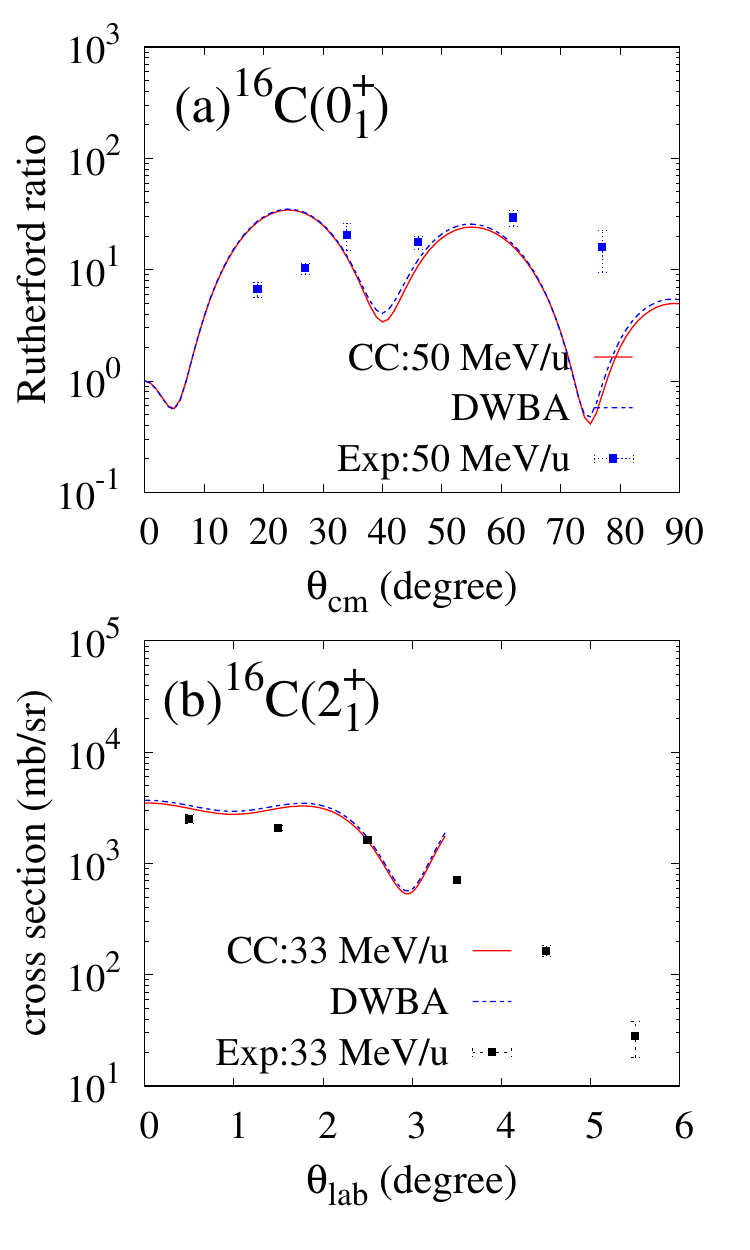}
  \caption{(a) Cross sections of the elastic proton scattering off $^{16}\textrm{C}$ at $E=50$ MeV/u and 
(b) those of the inelastic scattering at $E=33$ MeV/u
calculated by the CC calculation with the AMD densities (red solid lines).
The one-step cross sections obtained by the DWBA calculation are also shown (blue dotted lines).
The calculations are compared with the experimental data \cite{Grassi:2012kh,Ong:2006rm}
measured in inverse kinematics.
For the inelastic scattering,
$\theta_\textrm{lab}$ is kinematically limited within  3.6 degrees
but the data
contain effects of finite size and angular spread of the incident beam, multiple scattering in the target,
and detector geometry.
\label{fig:cross-c16}}
\end{figure}

The structure studies of $^{10}$Be, $^{12}$Be,  and $^{16}$C with AMD have been done
in Refs.~\cite{Kanada-Enyo:1999bsw,Kanada-Enyo:2003fhn,Kanada-Enyo:2004tao}.
We here briefly describe the structure properties, in particular, of the ground bands in these nuclei.

In $^{10}$Be,
the $M_n/M_p=0.9$ of the ground-band transition $2^+_1\to 0^+_1$ in the AMD calculation
 is smaller than $N/Z=1.5$ naively expected from the collective model and 
in reasonable agreement with the experimental value ($M_n/M_p=1.1$) reduced from $B(E2)$ of mirror transitions.
It indicates that the neutron excitation is somewhat suppressed 
compared with the proton excitation in the $2^+_1$ state.
In addition to the $2^+_1$ state of the ground $K^\pi=0^+$ band,
the $2^+_2$ state of the side band ($K^\pi=2^+$) is obtained because of the triaxial deformation.
In the higher energy region, the second $K^\pi=0^+$ band with the developed cluster structure appears.

In  $^{12}$Be,  the breaking of $N=8$ magicity is known in the ground state from
various experimental observations such as the Gamov-Tellar transitions, inelastic scattering,
and knock-out reactions
\cite{Iwasaki:2000gh,Suzuki:1997zza,Navin:2000zz,Pain:2005xw,Imai:2009zza,Meharchand:2012zz}.
The AMD calculation obtains the largely deformed ground band ($K^\pi=0^+_1$) with the dominant
neutron $2\hbar\omega$ component.
The ground-band transition, $2^+_1\to 0^+_1$, is strong because of the large deformation
compared with the weaker transition in the $K^\pi=0^+_2$ band, which corresponds to
the normal neutron $p$-shell closed  configuration.
In particular, the neutron transition is considerably strong
because of the contribution of two $sd$-orbit neutrons.
The values of the ratio $M_n/M_p=1.9$ ($M_n/M_p=1.1$) is obtained for the
$K^\pi=0^+_1$($K^\pi=0^+_2)$ band. The ratio of the ground band is as large as $N/Z=2$
because of the breaking of the $N=8$ magicity.

In the case of $^{16}$C, the AMD calculation predicted the
weak proton transition in  $2^+_1\to 0^+_1$
because of the $Z=6$ sub-shell closure.  The observed $B(E2)$ values are consistent with the
prediction. On the other hand, the neutron transition is significantly large
because of the contribution of the $sd$-orbit neutrons, and results in
the much larger ratio $M_n/M_p=3.2$ than $N/Z=1.67$, i.e.,
the dominant neutron contribution in the ground-band transition.

Figure \ref{fig:trans-be12} shows the neutron and proton matter densities of the ground
 state
and the neutron and proton  transition densities of the $2^+_1\to 0^+_1$ transition of
$^{10}$Be, $^{12}$Be, and $^{16}$C.
In $^{10}$Be,
the proton and neutron transition densities have the peak amplitude
at the same position at the nuclear surface
and approximately
satisfy the relation $\rho^\textrm{tr}_n(r)=(M_n/M_p) \rho^\textrm{tr}_p(r)$.
By contrast, in $^{12}$Be and $^{16}$C,
the neutron transition density shows the $r$ behavior quite different from the proton transition density.
It has the peak amplitude in
$r\approx 3$ fm region much outer than the proton transition density
because of the contribution of the $sd$-orbit neutrons
and no longer satisfies the relation $\rho^\textrm{tr}_n(r)=(M_n/M_p) \rho^\textrm{tr}_p(r)$.
This is a different feature from $^{10}$Be, where the protons and neutrons in the same $p$ shell
contribute to the $2^+_1$ excitation.

The proton scattering cross sections are calculated with the AMD densities.
For $^{10}$Be, the theoretical proton and neutron transition densities
are renormalized to fit the experimental transition strengths ($B^{(p),(n)}_{\lambda=2}$)
by the scaling factors
$M^\textrm{exp}_{p,n}/ M^\textrm{cal}_{p,n}$ listed in Table \ref{tab:radii}.
For $^{12}$Be and $^{16}$C, we use the original AMD transition densities, which reproduce well the
experimental $B(E2)$ values.
The calculated elastic and inelastic cross sections of $^{10}$Be at $E=60$ MeV/u, $^{12}$Be at $E=55$ MeV/u,
and $^{10}$C at $E=45$ MeV/u are shown in Figs.~\ref{fig:cross-be12-ela} and ~\ref{fig:cross-be12-inel}.
They are compared with the experimental data measured  in inverse kinematics.
In Fig.~\ref{fig:cross-c16}, the calculated cross sections of   $^{16}$C at $E=33$ MeV/u are compared with the experimental data.
The present calculation reproduces well the absolute amplitude of the $2^+_1$ cross sections as well as the
elastic cross sections.

In Fig.~\ref{fig:trans-be12}(b) for $^{12}$Be, 
the DWBA calculation with the neutron transition density $\rho^\textrm{tr}_n(r)=(M_n/M_p) \rho^\textrm{tr}_p(r)$
is also shown. This calculation corresponds to the case with the 
collective model transition density. Compared with the result using the original AMD transition density, 
the cross sections somewhat increase and the peak and dip positions slightly shift toward larger angles.

For the neutron transition in $^{12}$Be and $^{16}$C,  there is no data from the mirror nuclei.
The good reproduction of the inelastic cross sections supports reliability of the
neutron transition densities adopted in the present calculation, that is,
the dominant neutron contributions as $M_n/M_p\approx 2$ and $M_n/M_p\approx 3$ for
$^{12}$Be and $^{16}$C, respectively. This result is
qualitatively consistent with those in Refs.~\cite{Ong:2006rm,Takashina:2008zza}.
It should be stressed again that
phenomenological adjustable parameters were needed in
the reaction models of Refs.~\cite{Ong:2006rm,Takashina:2008zza}, but not in the present model.
For further detailed discussion of the transition densities,
higher quality data are required.

\subsection{Discussions}

\begin{figure}[!h]
\includegraphics[width=8 cm]{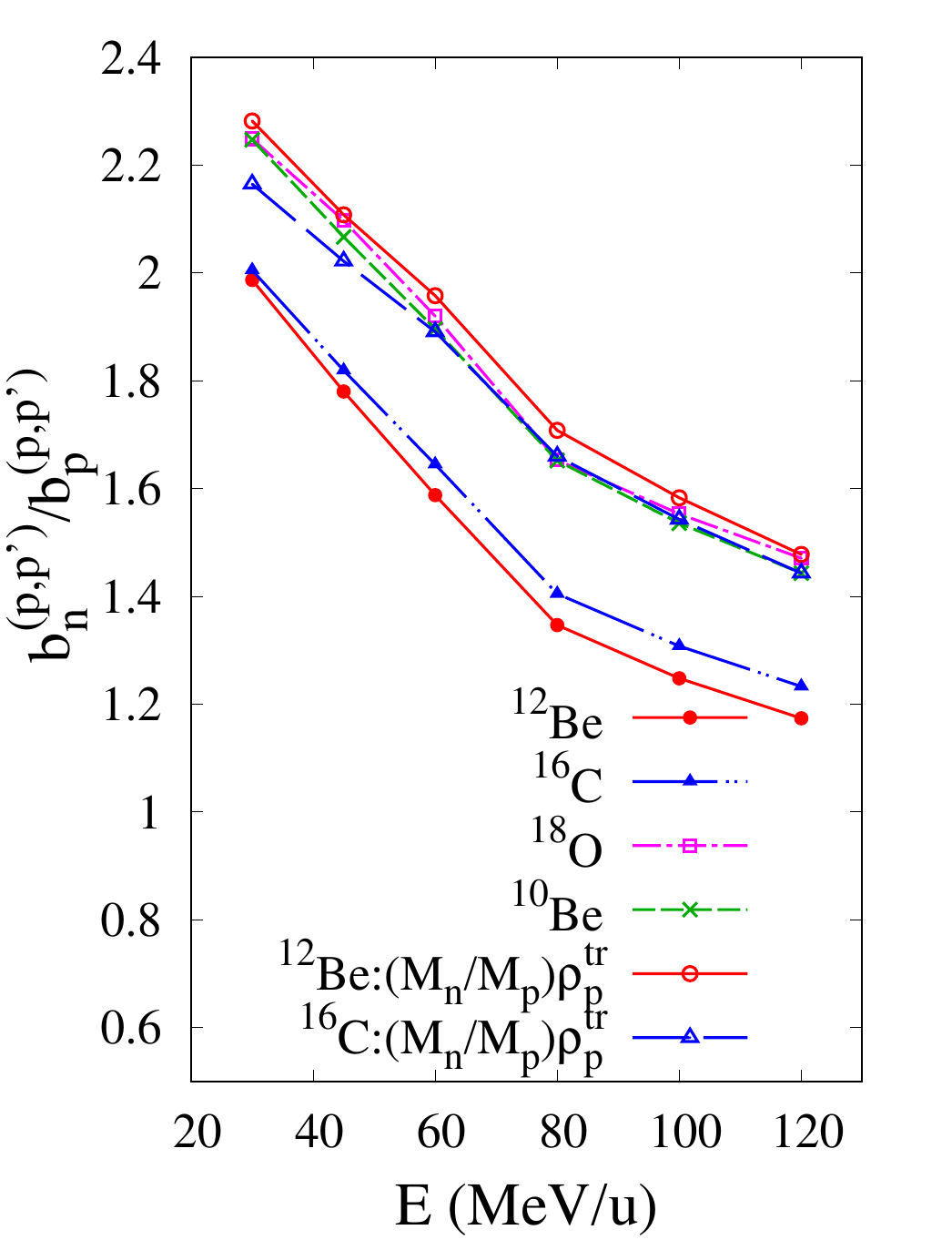}
  \caption{$b^{(p,p')}_n/b^{(p,p')}_p$ ratio of the proton scattering at $E=30$ MeV/u, 45 MeV/u, 60 MeV/u, 80 MeV/u, 100 MeV/u, and 120 MeV/u.
The values for $^{18}$O (magenta open squares), $^{10}$Be (green crosses), $^{12}$Be (red filled circles),
and $^{16}$C (blue filled triangles) calculated with the
default densities, 
and those for $^{12}$Be (red open circles), and $^{16}$C (blue open triangles)
of the $\rho^\textrm{tr}_n(r)=(M_n/M_p) \rho^\textrm{tr}_p(r)$ case are shown.
  \label{fig:bnbp}}
\end{figure}

We discuss how one can link the inelastic proton scattering cross sections
with the neutron transition matrix element $M_n$ of the $0^+_1\to 2^+_1$ transition.
The experimental studies of Refs.~\cite{Iwasaki:2000gh,Ong:2006rm} have discussed
the neutron matrix elements of $^{12}$Be and $^{16}$C with the reaction analysis of the
proton scattering data,
and concluded the significant neutron contribution in the $0^+_1\to 2^+_1$ transition.
According to the model analysis in Refs.~\cite{Iwasaki:2000gh,Ong:2006rm}
using the Bernstein prescription \cite{Bernstein:1981fp},
$B^{(n)}_{\lambda=2}=17$ fm$^4$
of $^{12}$Be is obtained by the reaction analysis using $B(E2)=14.2(2.8)$ $e^2$fm$^4$,
and $B^{(n)}_{\lambda=2}=25$ fm$^4$ of $^{16}$C is reduced using the updated data of
$B(E2)=2.6(9)$ $e^2$fm$^4$ \cite{Ong:2006rm}.
The value of $^{16}$C is consistent with our value of $B^{(n)}_{\lambda=2}=27.0$ fm$^4$, but the value of
$^{12}$Be is much smaller than $B^{(n)}_{\lambda=2}=51.1$ fm$^4$ of the present calculation.  In the theoretical study of the proton scattering of $^{12}$Be
with a MCC calculation using the same AMD densities \cite{Takashina:2008zza},  the slightly smaller value
$B^{(n)}_{\lambda=2}=37$ fm$^4$ was favored to reproduce the
inelastic cross sections.

The reaction analysis with the Bernstein prescription usually
assumes the simple collective model transition densities given by the derivative
of the matter density, and follows the
relation of inelastic hadron ($h,h'$) scattering cross sections with the
transition matrix elements as
\begin{align}\label{eq:bernstein}
\sigma(h,h') \propto  |b^{(h,h')}_n M_n+ b^{(h,h')}_p M_p|^2,
\end{align}
where $b^{(h,h')}_n$ and $b^{(h,h')}_p$ are the neutron and proton field strengths of the external field from the hadron probe.
For the proton scattering, $b^{(p,p')}_n/b^{(p,p')}_p$ depends on the
energy. A standard value $b^{(p,p')}_n/b^{(p,p')}_p\approx 3$ at $E=10$--50 MeV/u, which was obtained from the data of various ordinary nuclei, is often used.
The Bernstein prescription has been widely used for
the inelastic proton scattering, but it greatly relies on the
reaction model, which contains ambiguity
such as the proton-nucleus optical potentials and the
transition densities.
The ansatz of Eq.~\eqref{eq:bernstein} means the linear relation of
the squared cross section with the neutron ($M_n$) and proton  ($M_p$)
transition matrix elements. The ratio $b^{(p,p')}_n/b^{(p,p')}_p$
indicates the
sensitivity of the cross sections to the neutron transition matrix element ($M_n$) relative to the
proton part ($M_p$), and is supposed to be system independent.
The linear relation could be valid only
if the relation $\rho^\textrm{tr}_n(r)\propto \rho^\textrm{tr}_p(r)$ is satisfied.

However, it is not the case with $^{12}$Be and $^{16}$C, for which
the neutron transition density has the outer amplitude than the proton part.
Such the exotic behavior of the neutron transition density
may give non-trivial effects on the relation between
the cross sections and the transition matrix elements ($M_n$ and $M_p$).
To see this effect we microscopically derive the ratio $b^{(p,p')}_n/b^{(p,p')}_p$
within the present MCC approach and discuss how the sensitivity of the cross section to $M_n$ changes depending on the system
as well as on the incident energy.

Here we assume that the AMD calculation gives correct
$r$ dependence of the $\rho^\textrm{tr}_n(r)$ and $\rho^\textrm{tr}_p(r)$
but contains ambiguity of the overall factor in each of the neutron and proton parts.
By artificially changing the overall factor of $\rho^\textrm{tr}_n(r)$ or $\rho^\textrm{tr}_p(r)$, we calculate
the integrated cross sections and reduce the coefficients in the relation
\begin{align}
\sigma(p,p')=| a^{(p,p')}_n(^AZ,E)  M_n+ a^{(p,p')}_p(^AZ,E)  M_p|^2.
\end{align}
Here $a^{(p,p')}_n$ and $a^{(p,p')}_p$ are the system- and energy-dependent
parameters determined from the calculated cross sections.
The ratio $a^{(p,p')}_n/a^{(p,p')}_p$  is nothing but the ratio $b^{(p,p')}_n/b^{(p,p')}_p$ to be discussed.

In Fig.~\ref{fig:bnbp}, we show the obtained result of
$b^{(p,p')}_n/b^{(p,p')}_p=a^{(p,p')}_n/a^{(p,p')}_p$ for each systems,
$^{18}$O, $^{10}$Be, $^{12}$Be, and $^{16}$C, with the
default transition densities 
by open squares, crosses, filled circles, and filled triangles, respectively.
 For $^{12}$Be and $^{16}$C,
we also show the result for the naive case of $\rho^\textrm{tr}_n(r)\propto \rho^\textrm{tr}_p(r)$ by open circles and triangles, respectively, which are 
obtained by the MCC calculation using 
$\rho^\textrm{tr}_n(r)=(M_n/M_p) \rho^\textrm{tr}_p(r)$.
In all the cases, the calculated values of $b^{(p,p')}_n/b^{(p,p')}_p$ show
similar energy dependence, i.e.,
decreasing with the increase of $E$.
This energy dependence mainly comes from
the energy dependence of the effective $g$-matrix $NN$ interactions.
In fact, if the nucleon-nucleus optical potentials fixed at $E=60$ MeV/u is used,
$b^{(p,p')}_n/b^{(p,p')}_p$ varies slightly from 1.90 (1.64)  to 1.89 (1.57)
in the energy range  $30$--120~MeV/u for the proton scattering off
$^{10}$Be ($^{12}$Be).

At each energy, almost the same values of $b^{(p,p')}_n/b^{(p,p')}_p$ are obtained for
$^{18}$O and $^{10}$Be, and also by the calculation with 
the $\rho^\textrm{tr}_n(r)=(M_n/M_p) \rho^\textrm{tr}_p(r)$ assumption for  
$^{12}$Be and $^{16}$C.
These values can be regarded as standard values for ordinary systems with $\rho^\textrm{tr}_n(r)\approx (M_n/M_p) \rho^\textrm{tr}_p(r)$.
However, in the exotic case
$\rho^\textrm{tr}_n(r)\ne (M_n/M_p) \rho^\textrm{tr}_p(r)$ of
$^{12}$Be and $^{16}$C with the default transition densities,
the values of $b^{(p,p')}_n/b^{(p,p')}_p$ significantly deviate from the standard values:
they are smaller than the standard values by about 0.3 indicating that
the sensitivity of the cross sections to $M_n$ is by about $15\%$ weaker
than the ordinary case of $\rho^\textrm{tr}_n(r)=(M_n/M_p) \rho^\textrm{tr}_p(r)$.
The reason for the weaker sensitivity of the cross sections is that
the outer amplitude of the neutron transition density significantly
contributes to $M_n$ but does not so much to the cross sections.

This result may suggest a possible modification of the
phenomenological reaction analysis.
For simplicity, let us suppose that
there is no ambiguity in the reaction model
except for the neutron transition density $\rho^\textrm{tr}_n(r)$,
and other inputs are so reliable that
the model can properly reproduce the
cross sections for the ordinary case.
If one performs an inconsistent analysis using
$\rho^\textrm{tr}_n(r)=(M_n/M_p) \rho^\textrm{tr}_p(r)$ for the exotic case,
one could draw an underestimated value of $M_n$
from the observed cross sections.

\section{Summary}  \label{sec:summary}

We investigated
the proton inelastic scattering off $^{18}$O, $^{10}$Be, $^{12}$Be,  and $^{16}$C
to the $2^+_1$ states with the
microscopic coupled-channel calculation.
The proton-nucleus potentials are microscopically
derived by folding the Melbourne $g$-matrix $NN$ interaction with
the AMD densities of the target nuclei.
The calculated result reasonably reproduces the
elastic and inelastic proton scattering cross sections, and supports the dominant neutron contribution
in the $2^+_1$ excitation of $^{12}$Be and $^{16}$C.
In order to discuss the detailed behavior of transition densities, further high quality data of the differential
cross sections are required.

The sensitivity of the inelastic scattering cross sections to the neutron transition density was discussed.
A particular attention was paid on the exotic systems such as $^{12}$Be and $^{16}$C
that the neutron transition density
has the remarkable amplitude in the outer region than the proton part.
This outer amplitude of the neutron transition density significantly contributes to the
neutron matrix element
$M_n$. However, its contribution to the inelastic cross sections is quite modest because the reaction process considered has no strong selectivity for the outer region. 
This result indicates that a phenomenological analysis with the Bernstein prescription is no longer valid.
Our finding will suggest that a phenomenological analysis with collective model transition densities can result in an undershooting of $M_n$ for such exotic systems.

\begin{acknowledgments}
The computational calculations of this work were performed by using the
supercomputer in the Yukawa Institute for theoretical physics, Kyoto University. This work was partly supported
by Grants-in-Aid of the Japan Society for the Promotion of Science (Grant Nos. JP18K03617, JP16K05352, and 18H05407) and by the grant for the RCNP joint research project.
\end{acknowledgments}

\appendix

\section{Resummation factor in the folding model calculation}

According to the multiple scattering theory for nucleon-nucleus scattering constructed by Kerman, McManus, and Thaler~\cite{Ker59}, the transition matrix $T$ is given by
\begin{equation}
T=\frac{A}{A-1}T',
\label{tmat}
\end{equation}
where $A$ is the mass number of the nucleus and $T'$ is the transition matrix corresponding to the Schr{\"o}dinger equation
\begin{equation}
\bigg[K+h+\frac{A-1}{A}\sum_{j=1}^{A}\tau_{j}-E\bigg]\Psi=0.
\label{tau}
\end{equation}
$K$ is the kinetic energy operator, $h$ is the internal Hamiltonian of the nucleus, $E$ is the total energy, and $\Psi$ is the total wave function. $\tau_{j}$ is the effective interaction between the incident nucleon and a nucleon inside the nucleus, which is approximated to the Melbourne $g$-matrix $NN$ interaction in this study. The two factors, $A/(A-1)$ in Eq.~(\ref{tmat}) and $(A-1)/A$ in Eq.~(\ref{tau}) appear as a result of the {\it resummation} of the $NN$ collisions originally written in terms of a bare $NN$ interaction. Although these resummation factors usually do not play a role, for nucleon scattering off a light nucleus especially at low energies, these may slightly modify the result as shown in Ref.~\cite{Minomo:2017hjl}. 
These factors are taken into account in all the calculations shown in the present calculation.

\end{document}